# A charge density wave-like instability in a doped spin-orbit assisted weak Mott insulator

H. Chu[1,2], L. Zhao[2,3], A. de la Torre[2,3], T. Hogan[4], S. D. Wilson[4] & D. Hsieh[2,3]

[1]Department of Applied Physics, California Institute of Technology, Pasadena, CA 91125, USA.

[2]Institute for Quantum Information and Matter, California Institute of Technology, Pasadena, CA 91125, USA.

[3]Department of Physics, California Institute of Technology, Pasadena, CA 91125, USA.

[4]Materials Department, University of California, Santa Barbara, CA 93106, USA.

**Layered perovskite iridates realize a rare class of Mott insulators that are predicted to be strongly spin-orbit coupled analogues of the parent state of cuprate high-temperature superconductors[1,2]. Recent discoveries of pseudogap[3–5], magnetic multipolar ordered[6] and possible *d*-wave superconducting phases[7,8] in doped $Sr_2IrO_4$ have reinforced this analogy among the single layer variants. However, unlike the bilayer cuprates[9], no electronic instabilities have been reported in the doped bilayer iridate $Sr_3Ir_2O_7$. Here we show that $Sr_3Ir_2O_7$ realizes a weak Mott state with no cuprate analogue[9] by using ultrafast time-resolved optical reflectivity to uncover an intimate connection between its insulating gap and antiferromagnetism. However, we detect a subtle charge density wave-like Fermi surface instability in metallic electron doped $Sr_3Ir_2O_7$ at temperatures ($T_{DW}$) close to 200 K via the coherent oscillations of its collective modes, which is reminiscent of that observed in cuprates[10,11]. The absence of any signatures of a new spatial periodicity below $T_{DW}$ from diffraction[12], scanning tunneling[12] and photoemission[13,14] based probes suggests an unconventional and possibly short-ranged nature of this density wave order.**

The ground state of $Sr_3Ir_2O_7$ is an antiferromagnetic insulator with a charge gap ($\Delta$) of order 100 meV as determined by optical conductivity[15], scanning tunneling microscopy[16] (STM) and angle-resolved photoemission spectroscopy[17] (ARPES) measurements. A starting point for understanding this state is the spin-orbit Mott insulator model[1] in which the low energy electronic structure is approximated by half-filled pseudospin $J_{eff} = 1/2$ bands and filled $J_{eff} = 3/2$ bands that are derived from spin-orbit split Ir $t_{2g}$ orbitals. In the presence of on-site Coulomb repulsion, a Mott gap is driven in the $J_{eff} = 1/2$ band without requiring antiferromagnetic ordering of the $J_{eff} = ½$ moments. However, experimental evidence suggests that $Sr_3Ir_2O_7$ does not adhere to this strong Mott insulator description. Notably, there is evidence from transport[18], optical conductivity[15] and ARPES[19] studies that the electronic density of states near the Fermi level changes in the vicinity of the antiferromagnetic ordering temperature ($T_N \sim 280$ K). Moreover, resonant inelastic x-ray scattering measurements[20,21] show that the energy scale of magnetic exchange is comparable to $\Delta$.

To identify the relationship between the insulating gap and antiferromagnetic ordering in Sr$_3$Ir$_2$O$_7$, we performed pump-probe optical reflectivity experiments in which an ultrashort optical (pump) pulse is used to excite electron-hole (e-h) pairs across the insulating gap and another time-delayed (probe) pulse is used to measure the fractional change in reflectivity ($\Delta R/R$) as a function of time delay (*t*) after the pump excitation. Such reflectivity transients track the relaxation dynamics of photo-excited carriers back to equilibrium and are known to be highly sensitive to even very small charge gaps[22,23] that may not be resolvable by ARPES or optical conductivity measurements especially at high temperatures. Further technical details of the experiment are provided in the Methods.

Figure 1a shows the reflectivity transients of Sr$_3$Ir$_2$O$_7$ measured using a pump and probe photon energy of 1.55 eV >> $\Delta$ plotted as a function of temperature. At all temperatures, the pump excitation causes an abrupt drop in reflectivity followed by a slower recovery on the picosecond timescale. It is clear from the data that both the magnitude of the drop and the recovery time become strongly temperature dependent around $T_N$. To understand the temperature dependence in greater detail, the individual reflectivity transients were fit to an exponential decay function $\Delta R/R = Ae^{-t/\tau} + C$ (see Supplementary Section S1 for individual fits) where *A* is the amplitude of the reflectivity drop at *t* = 0, $\tau$ is the characteristic photo-carrier relaxation time and *C* describes much slower equilibration processes such as heat diffusion. Figure 1b shows that upon heating towards $T_N$, *A* sharply declines in magnitude while $\tau$ exhibits a divergent behaviour, both of which are consistent with the insulating gap closing at $T_N$. Microscopically, this can be understood using the phenomenological Rothwarf-Taylor model[24] where *A* is treated as being representative of the fractional change in the number of e-h pairs introduced by the pump pulse and $\tau$ is the characteristic time for the population of such pairs to decay. Recombination of an e-h pair across the gap is accompanied by the emission of a phonon with energy greater than or equal to $\Delta$. Since these phonons can in turn excite additional e-h pairs, $\tau$ is ultimately determined by the time required for these phonons to anharmonically decay into phonons of energy less than $\Delta$ or to escape from the probed region. As the gap closes upon approaching $T_N$, more thermally populated low energy phonons become available for e-h pair production, impeding the recombination of photo-excited e-h pairs and reducing their ratio to thermally excited e-h pairs, which respectively cause $\tau$ to diverge and $A$ to precipitously drop. In the limit where the ratio of photo-excited to thermally excited e-h pairs is small, analytical expressions can be derived[25] for the temperature dependence of both $A$ and $\tau$ assuming a Bardeen-Cooper-Schrieffer (BCS)-like temperature dependence of the gap below $T_N$ (see Supplementary Section S2 for detailed expressions), which fit the data remarkably well as shown in Fig. 1b. These results confirm that the insulating state of $Sr_3Ir_2O_7$ is only marginally stable and that the insulating gap closes concurrently with the loss of antiferromagnetic long-range order, which differs from the strongly correlated Mott state realized in cuprates[9]. The fact that no drastic changes in band structure are observed across $T_N$ by ARPES[19] and optical conductivity[15] measurements suggests that the gap closure in $Sr_3Ir_2O_7$ primarily occurs through the transfer of high energy spectral weight into the gap (Fig. 1b inset) and is concentrated at momentum values near the X points of the Brillouin zone as shown by ARPES[19].

It has recently been shown that electron doping $Sr_3Ir_2O_7$ by substituting Sr with La can also suppress antiferromagnetism and force the insulating gap to close[12–14,26]. However unlike in the thermally driven case (Fig. 1), ARPES studies have shown that doping across the critical value of $x$ in $(Sr_{1-x}La_x)_3Ir_2O_7$, which is on the few percent level, causes a discontinuous shift of the band edges and the formation of electron pockets around the M points of the Brillouin zone[13,14]. The contrast between these two mechanisms of gap closing is highly apparent from ultrafast optical reflectivity data. As shown in Fig. 2, the doping induced gap closing across $x_c \sim 0.02$ (2 %) is manifested as a change in sign of $\Delta R/R$, which is consistent with an insulator-to-metal transition that is realized by a collective electronic rearrangement[27]. Interestingly, across $x_c \sim 2$ %, an abrupt change is also observed in the collective excitations of the system. For $x < 2$ %, we observe a high frequency modulation superposed atop the exponential decay (Fig. 2 inset), which arises from coherent oscillations of a 4.4 THz $A_{1g}$ phonon mode that has also been observed by Raman scattering in $Sr_3Ir_2O_7$[28].

For all $x > 2$ % on the other hand, an additional low frequency (~ 1 THz) modulation is present (Fig. 2 inset) that does not correspond to any reported Raman active phonon mode. Unlike the $A_{1g}$ phonon mode, which diminishes in amplitude upon entering the metallic regime due likely to screening by free carriers, this mode only appears in the metallic regime suggesting an electronic origin.

To understand the origin of this low frequency mode, we studied its detailed temperature dependence. Figure 3a shows the temperature evolution of the reflectivity transients obtained from the most metallic sample ($x \sim 5.8$ %) available. At low temperatures, both the low frequency mode and the $A_{1g}$ phonon mode are clearly resolved in both the raw data and its Fourier transform (Fig. 3a inset). As the temperature increases, the frequency and the amplitude of the low frequency mode continuously decrease until the mode is no longer resolvable near 200 K. To extract quantitative information from this data, we fit the oscillatory component of the reflectivity transients to a damped sinusoid of the form $A_{DW} e^{-\Gamma t} \sin(2\pi t/\tau_{DW} + \phi)$ where $A_{DW}$, $\Gamma$ and $\tau_{DW}$ are the amplitude, damping rate and period of the low frequency mode, respectively, and $\phi$ is a phase offset (see Supplementary Section S3 for detailed fits). Figure 3b shows that $A_{DW}$ exhibits an order parameter like onset below a critical temperature $T_{DW} \sim 210$ K and Fig. 3c shows that both $\Gamma$ and $\tau_{DW}$ diverge upon approaching $T_{DW}$ indicating a softening of the mode to zero energy. These behaviours (together with the pump fluence dependence of $\tau_{DW}$ shown in Supplementary Section S4) are consistent with coherent amplitude oscillations of an electronic order parameter that develops via a continuous phase transition below $T_{DW}$. In particular, they are highly reminiscent of the coherent charge density wave amplitudons that have previously been observed by time-resolved optical reflectivity in systems such as $K_{0.3}MoO_3$[29] or the under-doped cuprates[10,11].

A recent neutron diffraction experiment found evidence of a weak intra-unit-cell structural distortion in metallic $(Sr_{1-x}La_x)_3Ir_2O_7$ crystals below a critical temperature $T_s$ that is also of order 200 K[12]. No new spatial periodicity was reported below $T_s$, which is consistent with the absence of additional superlattice modulations and band folding in low temperature

STM[12] and ARPES[13,14] data on $(Sr_{1-x}La_x)_3Ir_2O_7$ respectively. Based on the observations that the structural distortion emerges upon very light La doping, appears to compete with antiferromagnetic ordering and is suppressed in temperature by isovalent substitution of $Sr^{2+}$ by $Ca^{2+}$, it was suggested that there is a strong electronic contribution to the lattice distortion that enhances $T_s$[12]. By carrying out transient reflectivity measurements analogous to Fig. 3a-c for a series of different La doping levels (see Supplementary Section S5 for detailed comparison), we identified the region of the temperature versus doping phase diagram over which our observed electronic order exists (Fig. 3d). Remarkably, its phase boundary coincides well with that of the structural transition, reaffirming a cooperative interaction between the structural and electronic order parameters.

To confirm that the low frequency oscillations observed in $(Sr_{1-x}La_x)_3Ir_2O_7$ originate from a Fermi surface instability and not simply from the structural distortion below $T_s$, we performed time-resolved optical reflectivity measurements on isovalent substituted $(Sr_{1-x}Ca_x)_3Ir_2O_7$, which remains insulating like $Sr_3Ir_2O_7$ but undergoes the same structural distortion as that observed in metallic $(Sr_{1-x}La_x)_3Ir_2O_7$, albeit at considerably lower temperature (see Supplementary Section S6). Figure 4 shows the reflectivity transients of $(Sr_{0.93}Ca_{0.07})_3Ir_2O_7$ taken both above and below its structural transition temperature ($T_s \sim 120$ K). The $A_{1g}$ phonon mode at 4.4 THz is again observed and does not undergo any measurable shift across $T_s$. However, the low frequency mode that was observed in metallic $(Sr_{1-x}La_x)_3Ir_2O_7$ is clearly absent both in the raw data and its Fourier transform (Fig. 4 inset), thus precluding a purely structural origin of the mode.

No static long-range magnetic order has been observed by neutron diffraction, resonant x-ray diffraction or magnetic susceptibility measurements in metallic $(Sr_{1-x}La_x)_3Ir_2O_7$ crystals[12]. In addition, we have performed resonant inelastic x-ray scattering measurements (see Ref. [30] and Supplementary Section S7) that show both quasi two-dimensional magnon modes and exciton-like spin-orbital entangled modes in $(Sr_{1-x}La_x)_3Ir_2O_7$ to exist only at frequencies far above 1 THz. These observations rule out any spin-related instabilities at the

1 THz energy scale and instead point towards a charge density wave instability as being responsible for the low energy mode. However, this interpretation must be reconciled with the absence of any signatures of a new spatial periodicity below $T_{DW}$ in other experiments. One possibility is that the charge density wave order is very short-ranged like in the case of under-doped cuprates[10,11] or even temporally fluctuating, which would naturally explain the strong damping of its collective modes (Fig. 3c) in comparison to those in static long-range ordered charge density wave systems like $K_{0.3}MoO_3$[29]. While a detailed microscopic understanding of this density wave instability in $(Sr_{1-x}La_x)_3Ir_2O_7$ will require more extensive theory and experiments, our work shows generally that iridates in the weak Mott limit can also support interesting electronic instabilities, possibly realizing strongly spin-orbit coupled analogues of those found in their lighter transition metal oxide counterparts.

**Acknowledgements**

We thank V. Madhavan and Z. Wang for providing scanning tunnelling microscopy data and analysis on $(Sr_{1-x}La_x)_3Ir_2O_7$ samples and for helpful discussions. This work is supported by a GIST-Caltech Collaboration Grant and by ARO Grant W911NF-13-1-0059. Instrumentation was partially supported by ARO DURIP Award W911NF-13-1-0293. D.H. acknowledges funding provided by the Institute for Quantum Information and Matter, an NSF Physics Frontiers Center (PHY-1125565) with support of the Gordon and Betty Moore Foundation through Grant GBMF1250. S.D.W. acknowledges support under NSF award No. DMR-1505549 as well as partial support from the MRSEC Program of the National Science Foundation under Award No. DMR 1121053 (T. H.).

**Author contributions**

H.C., L.Z. and D.H. planned the experiment. H.C. performed the measurements. H.C., L.Z., A.d.l.T. and D.H. analyzed the data. T.H. and S.D.W. prepared and characterized the samples and performed RIXS measurements. H.C. and D.H. wrote the manuscript.

**Additional information**

Supplementary information is available in the online version of the paper. Correspondence and requests for materials should be addressed to D.H. (dhsieh@caltech.edu).

## Figure 1

**Temperature dependent reflectivity transients of $Sr_3Ir_2O_7$ ($x$ = 0). a,** Three-dimensional surface plot of the transient reflectivity as a function of temperature and delay measured using a pump fluence of 80 $\mu$J/cm$^2$. Black lines are raw traces at select temperatures. Pronounced changes are observed around $T_N$ ~ 280 K. **b,** Temperature dependence of the amplitude $A$ (open orange circles) and decay time $\tau$ (filled blue circles) of the reflectivity transients extracted from fits to the single exponential decay function. Error bars represent the standard deviation of the best fit parameters across 20 independent measurements made at each temperature and can largely be attributed to laser intensity fluctuations. The superposed orange and blue curves are best fits to the Rothwarf-Taylor model described in the main text. Inset depicts the transfer of spectral weight into the insulating gap above $T_N$ while the $J_{eff}$ = 1/2 valence band (VB) and conduction band (CB) remain intact.

## Figure 2

**Doping dependent reflectivity transients of $(Sr_{1-x}La_x)_3Ir_2O_7$.** Three-dimensional surface plot of the transient reflectivity as a function of La doping ($x$) and delay. Black lines are the raw traces of the $x$ = 0, 1.3, 2.7, 3.6, 4.7 and 5.8 % doped samples. All traces were measured at 120 K using a pump fluence of 140 $\mu$J/cm$^2$. Top inset shows a magnified view of the delay range bounded by triangles showing the high frequency coherent $A_{1g}$ phonon mode and the low frequency electronic mode observed in the $x$ = 1.3 % and $x$ = 2.7 % samples, respectively. The evolution of the band structure across the insulator-to-metal transition ($x_c$ ~ 2 %) along the M-X-M momentum space cut as determined by ARPES[13,14] is illustrated schematically below the data.

## Figure 3

**Coherent oscillations of an electronic order parameter. a,** Reflectivity transients of 5.8 % La doped $Sr_3Ir_2O_7$ plotted for a series of temperatures separated by 20 K. The traces are vertically offset for clarity. A higher pump fluence of 400 $\mu$J/cm$^2$ was used in order to better visualize the weak temporal modulations but they are present even at lower fluences (see Fig. 2). Two of the modulation peaks are marked with triangles to guide the eye. The Fourier transform of the background subtracted reflectivity transient measured at 180 K is shown in the inset. The red line is a best fit to the sum of two Lorentzian functions and a constant background. **b,c** Temperature dependence of the oscillation amplitude ($A_{DW}$), period ($\tau_{DW}$) and damping time ($\Gamma^{-1}$) extracted from fits to a damped sinusoidal function described in the main text. Error bars represent the 95% confidence intervals of the fitted values. **d,** Temperature versus doping phase diagram of $(Sr_{1-x}La_x)_3Ir_2O_7$. The paramagnetic (PM) to antiferromagnetic (AFM) phase boundary and the line of structural phase transitions ($T_s$) are based on data from Ref. [12] whereas the line of density wave phase transitions ($T_{DW}$) is based on time-resolved optical reflectivity data. Error bars denote the uncertainty in $T_{DW}$ estimated by extrapolating to the temperature at which $A_{DW}$ goes to zero.

## Figure 4

**Reflectivity transients of insulating $(Sr_{1-x}Ca_x)_3Ir_2O_7$.** Transient reflectivity of ~ 7 % Ca substituted $Sr_3Ir_2O_7$ plotted over a range of temperatures spanning its structural transition temperature $T_s$ ~ 120 K (see Supplementary Section S6). The traces are vertically offset for clarity. The $A_{1g}$ phonon oscillations are clearly resolved in both the raw data and Fourier transform (inset) but the electronic mode is absent. The red line in the inset is a best fit to the sum of a Lorentzian function and a constant background.

## Methods

**Material growth.** Single crystals were synthesized via a flux growth technique. High purity powders of $SrCO_3$, $IrO_2$, and $La_2O_3$ (Alfa Aesar) were dried, and stoichiometric amounts were measured out, employing a 15:1 molar ratio between $SrCl_2$ flux and target composition. Powders were loaded into a platinum crucible and further contained inside alumina crucibles to limit volatility. Mixtures were heated to 1300 °C, slowly cooled to 850 °C at a rate of 3.5 °C/hr, and then furnace cooled to room temperature. The resulting boule was etched with deionized water revealing black plate-like $(Sr_{1-x}La_x)_3Ir_2O_7$ crystals with typical dimensions 2 mm x 2 mm x 0.1 mm. Details of doping level and structural and electronic properties characterization can be found in Ref. [12].

**Time-resolved reflectivity measurements.** Time-resolved optical pump-probe reflectivity measurements were carried out using optical pulses of 80 fs temporal duration and 795 nm centre wavelength produced by a regeneratively amplified Ti:sapphire laser system operating at a 10 kHz repetition rate. The pump and probe pulses were cross polarized and focused onto the sample at near normal incidence. The pump induced change in the reflected probe beam intensity was measured using a lock-in amplifier referenced to the mechanical chopping frequency (5 kHz) of the pump beam. Use of a balanced detection scheme provided a sensitivity to fractional changes in reflectivity $\Delta R/R$ close to $10^{-5}$. The (001) surfaces of the crystals were cleaved in air prior to measurements and immediately pumped down to pressures better than $2 \times 10^{-7}$ mbar. All data presented in the main text were measured with a pump fluence less than 400 μJ/cm$^2$ and a probe fluence of 10 μJ/cm$^2$.

**Model of e-h pair relaxation dynamics.** In the limit where the population of photo-excited e-h pairs is much less than that of thermally excited e-h pairs, as shown in Ref. [25] the temperature dependence of the amplitude $A(T)$ and decay time $\tau(T)$ of the reflectivity transients ($\Delta R/R$) can be expressed as:

$$A(T) \propto \frac{\frac{F}{\Delta(T) + \frac{k_B T}{2}}}{1 + \gamma \sqrt{\frac{2k_B T}{\pi \Delta(T)}} e^{-\frac{\Delta(T)}{k_B T}}} \quad ; \quad \tau(T) \propto \frac{1}{\Delta(T)}$$

where $F$ is the pump fluence, $\gamma$ is a phenomenological fitting parameter and $k_B$ is Boltzmann's constant (see Supplementary Section S2 for details). To fit the data in Fig. 1b, we assumed that the insulating gap $\Delta(T)$ has a BCS like temperature dependence of the form $\Delta(T) = \Delta_0 \sqrt{1 - T/T_N}$ where both the gap at zero temperature ($\Delta_0$) and the antiferromagnetic ordering temperature ($T_N$) are fitting parameters.

**Data availability.** The data that support the plots within this paper and other findings of this study are available from the corresponding author upon reasonable request.

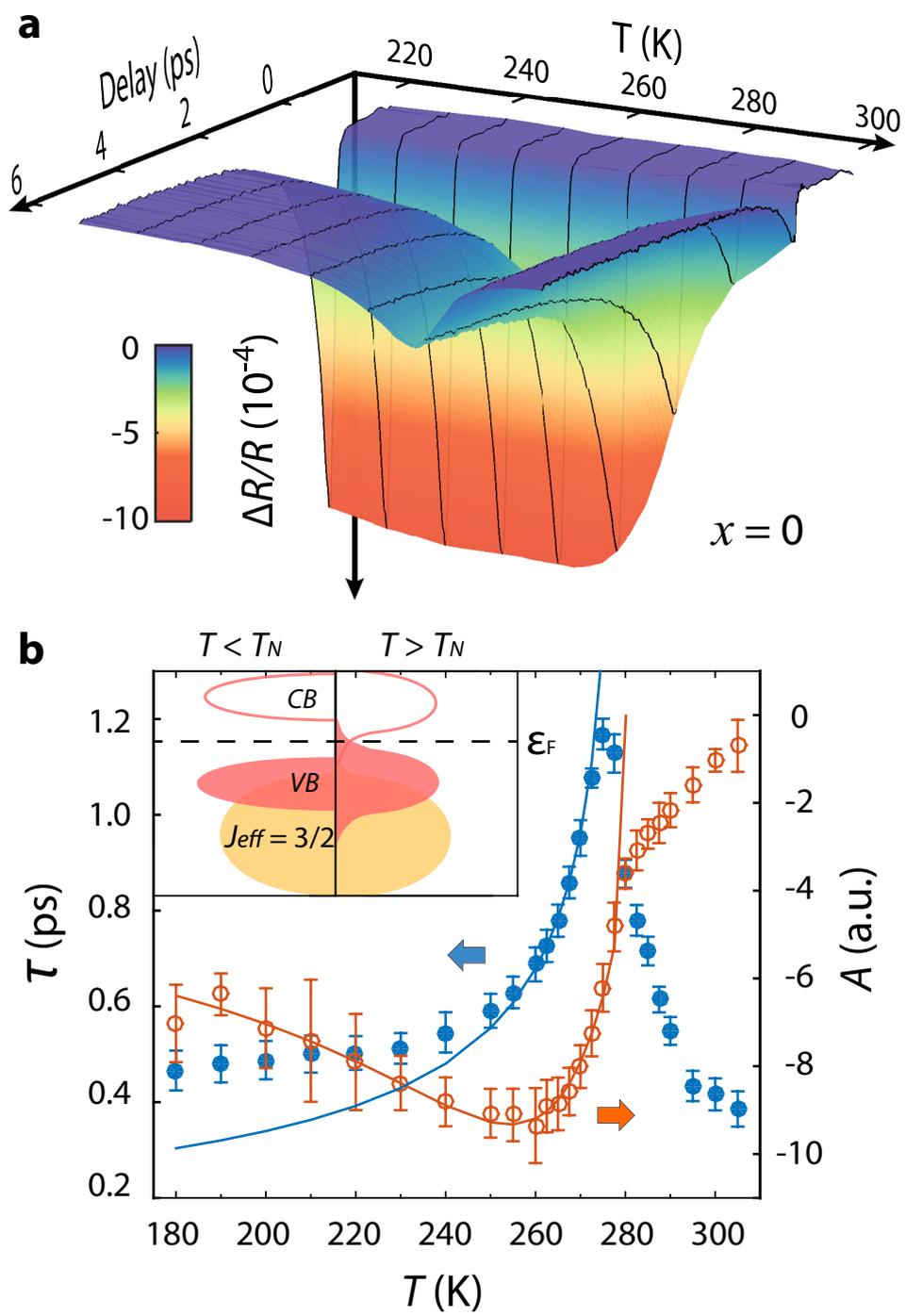

Figure 1

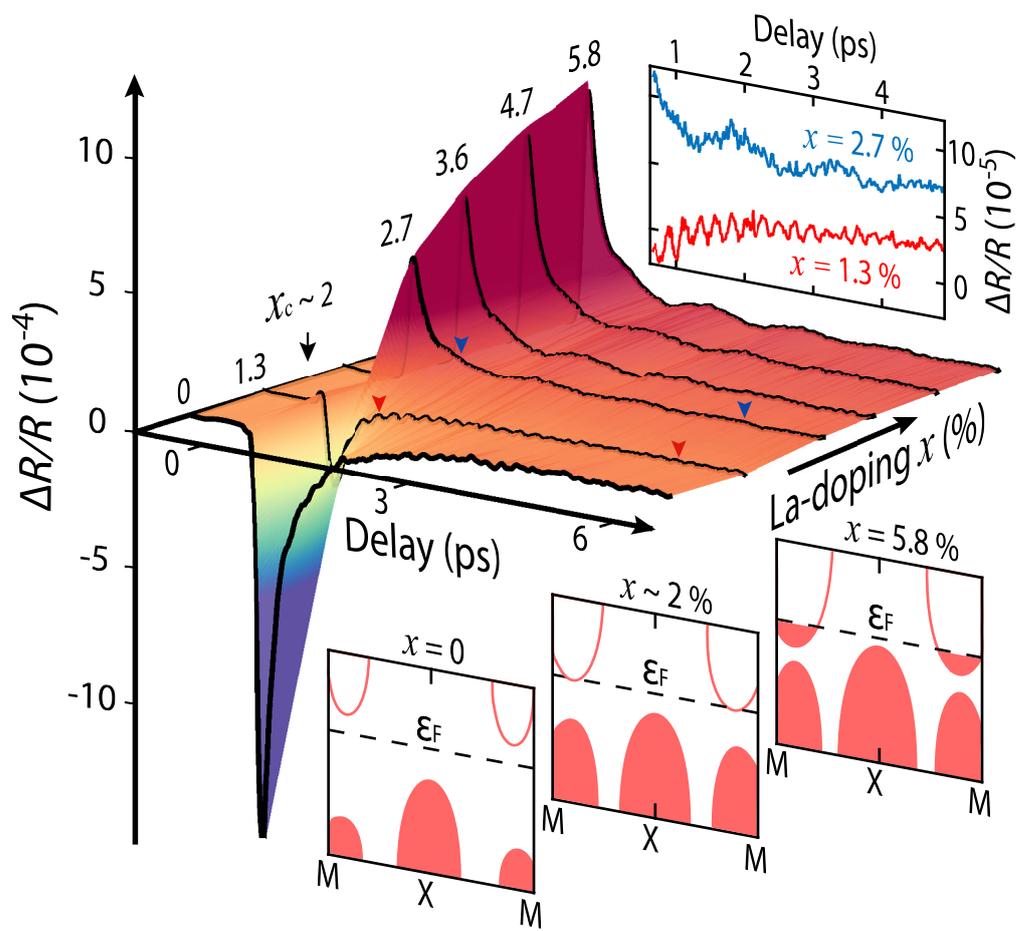

Figure 2

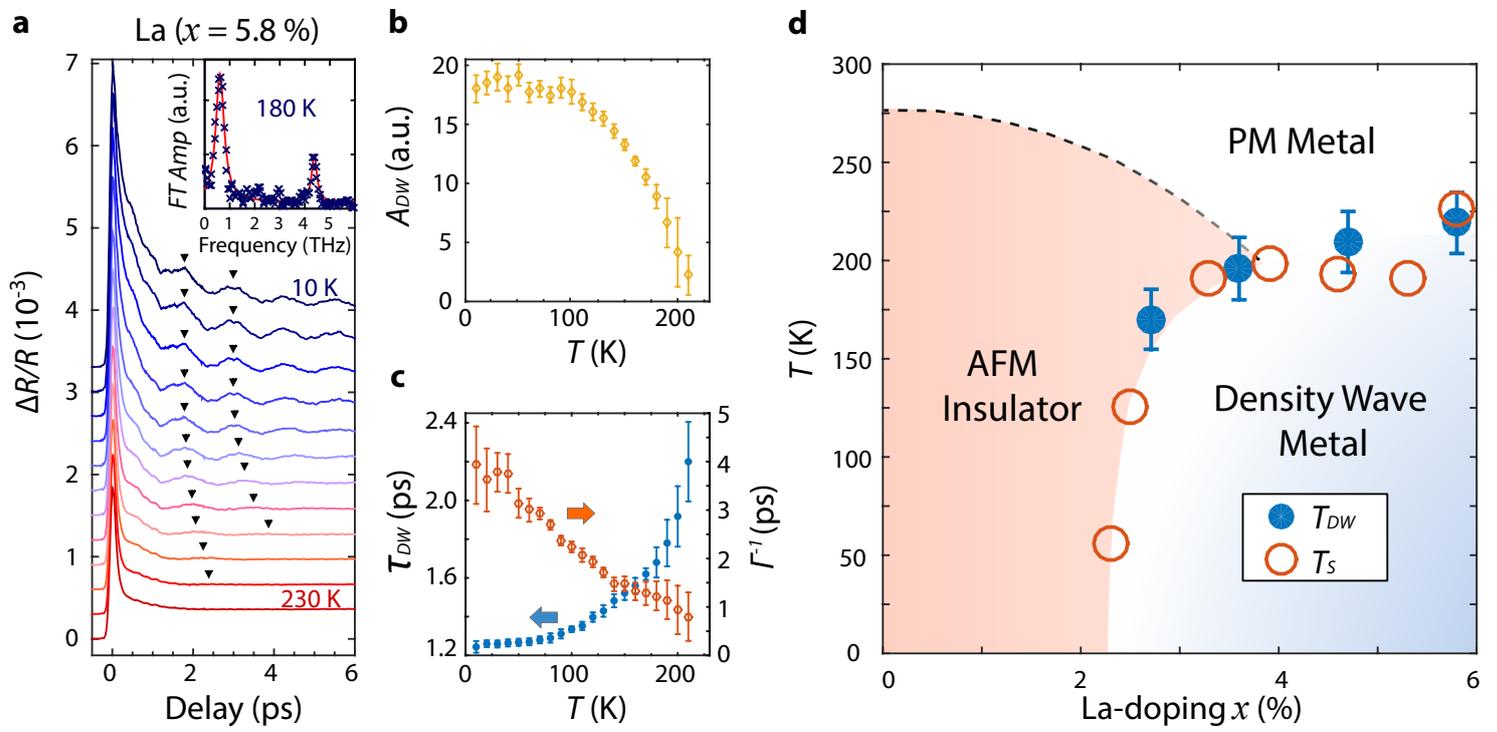

Figure 3

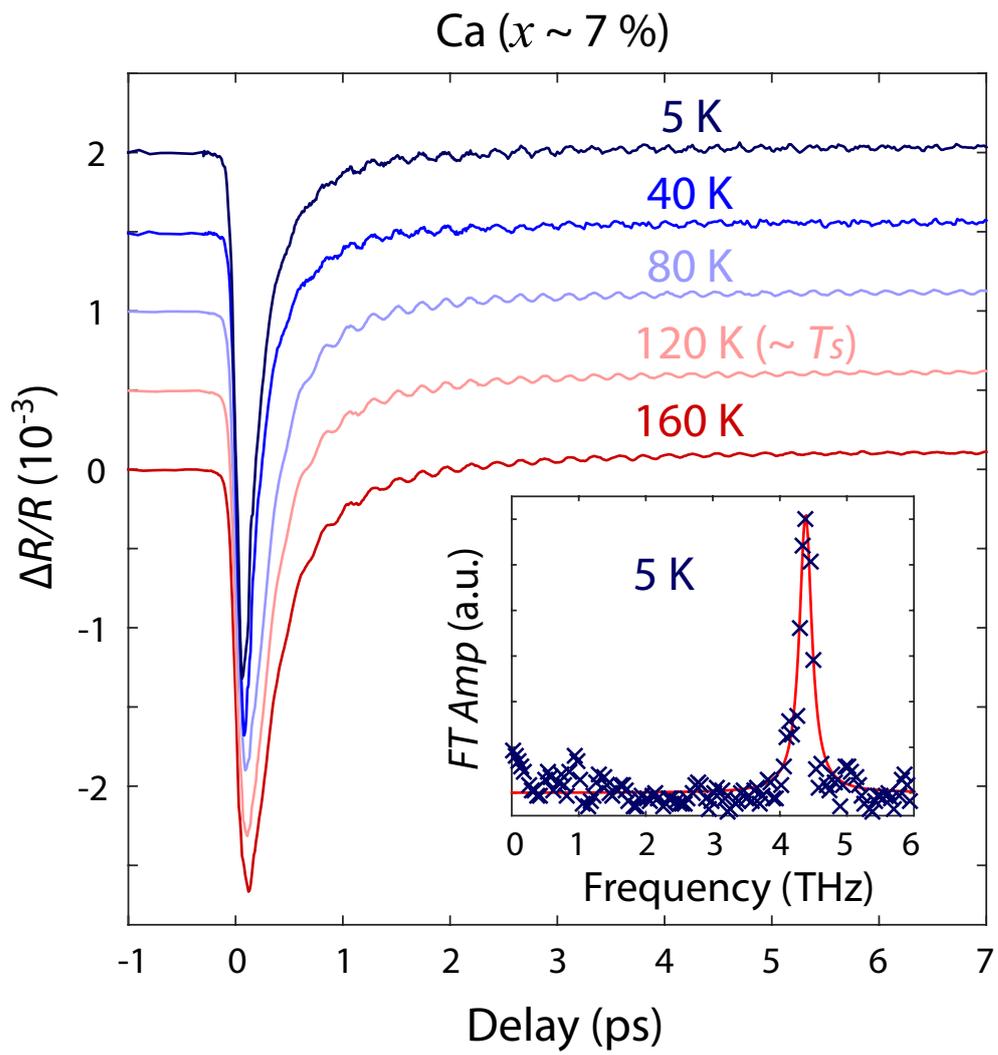

Figure 4

# Supplemental Information:

# A charge density wave-like instability in a doped spin-orbit assisted weak Mott insulator

H. Chu[1,2], L. Zhao[2,3], A. de la Torre[2,3], T. Hogan[4], S.D. Wilson[4] & D. Hsieh[2,3]

[1]Department of Applied Physics, California Institute of Technology, Pasadena, CA 91125, USA

[2]Institute for Quantum Information and Matter, California Institute of Technology, Pasadena, CA 91125, USA

[3]Department of Physics, California Institute of Technology, Pasadena, CA 91125, USA

[4]Materials Department, University of California, Santa Barbara, CA 93106, USA

S1. Fits of $Sr_3Ir_2O_7$ reflectivity transients to a single exponential decay

S2. Fits of $A(T)$ and $\tau(T)$ to photo-excited e-h pair relaxation model

S3. Fits of $(Sr_{1-x}La_x)_3Ir_2O_7$ reflectivity transients to a damped sinusoid

S4. Fluence dependence of electronic mode in $(Sr_{1-x}La_x)_3Ir_2O_7$

S5. La-doping dependence of $\tau_{DW}$

S6. Characterization of structural distortion and $T_s$ in $(Sr_{0.93}Ca_{0.07})_3Ir_2O_7$

S7. Magnon and spin-orbit exciton mode energies in $(Sr_{1-x}La_x)_3Ir_2O_7$

## S1. Fits of $Sr_3Ir_2O_7$ reflectivity transients to a single exponential decay

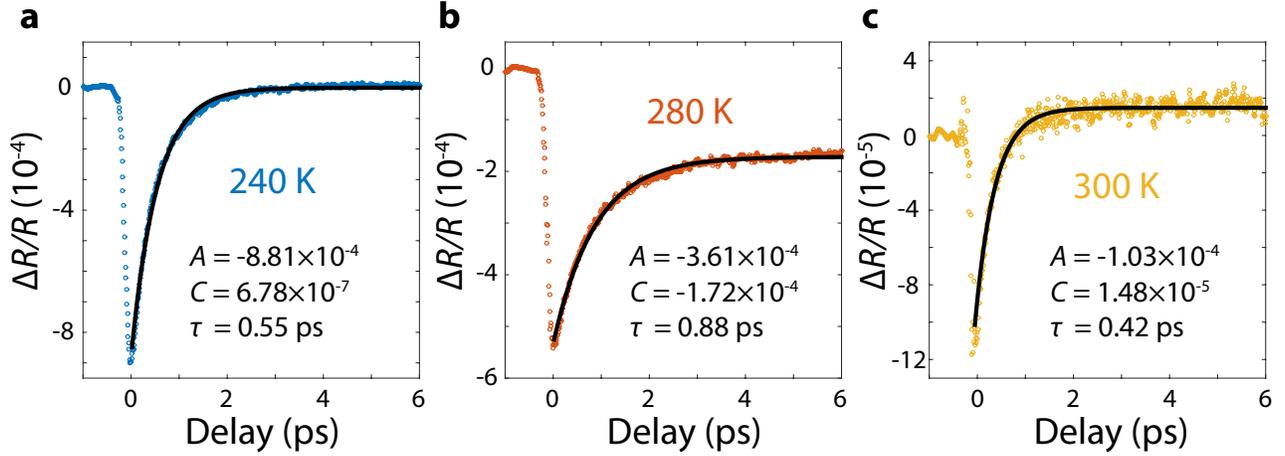

FIG. S1. Representative best fits of the reflectivity transients (open circles) of $Sr_3Ir_2O_7$ to a single exponential decay (solid lines) at temperatures **a,** below ($T = 240$ K), **b,** at ($T = 280$ K) and **c,** above ($T = 300$ K) the Néel temperature. The values of the best fit parameters are displayed in each panel.

The reflectivity transients of $Sr_3Ir_2O_7$ at all temperatures shown in Fig. 1a of the main text are well described by a single exponential decay function $\Delta R/R = Ae^{-t/\tau} + C$, where $A$ is the amplitude of the pump induced fractional change in reflectivity and $\tau$ is the characteristic electron-phonon relaxation time. Relaxation processes that are much slower than $\tau$, such as heat diffusion away from the probed region, are approximated as being constant ($C$) over the time interval of interest. Fits were constrained between delay times of 0.1 ps and 6 ps for all temperatures and the typical quality of fit is shown in Fig. S1.

**S2. Fits of $A(T)$ and $\tau(T)$ to photo-excited e-h pair relaxation model**

Analytical expressions for the temperature dependence of the amplitude $A(T)$ and relaxation time $\tau(T)$ of the reflectivity transients of a fully gapped (nodeless) system below its gap opening temperature were derived in Ref.[1] and are shown below.

$$A(T) \propto \frac{\frac{F}{\Delta(T) + \frac{k_B T}{2}}}{1 + \gamma \sqrt{\frac{2 k_B T}{\pi \Delta(T)}} e^{-\frac{\Delta(T)}{k_B T}}}$$

$$\tau(T) \propto \frac{1}{\Delta(T)}$$

Here $F$ is the fluence of the pump pulse, $k_B$ is Boltzmann's constant and $\gamma$ is a phenomenological fitting parameter. Although the derivation was made in the context of the recombination of quasiparticles across a superconducting gap, the same formalism extends to and is often used to describe the recombination of electron-hole pairs across an insulating gap.

To fit the data shown in Fig. 1b of the main text, we used a value $F = 80$ $\mu$J/cm$^2$ and assumed a BCS-like temperature dependence of the insulating gap $\Delta(T) = \Delta_0 \sqrt{1 - T/T_N}$. The zero temperature gap magnitude $\Delta_0$, Néel temperature $T_N$ and $\gamma$ were left as free parameters. Fits were constrained between temperatures of 180 K and 275 K. The best fits shown in Fig. 1b of the main text yielded the values $T_N = 280$ K, $\Delta_0 = 230$ meV and $\gamma = 12.2$. We note that this simple model produces the correct value of $T_N$ but slightly overestimates the value of $\Delta_0$.

## S3. Fits of $(Sr_{1-x}La_x)_3Ir_2O_7$ reflectivity transients to a damped sinusoid

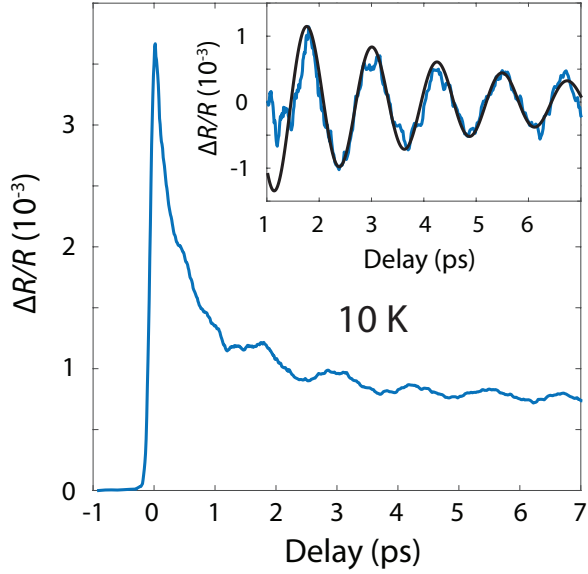

FIG. S2. Reflectivity transient of $(Sr_{1-x}La_x)_3Ir_2O_7$ with $x = 5.8\%$ measured at 10 K reproduced from Fig. 3a of the main text. Inset shows the background subtracted data (blue line) overlaid with a best fit to a damped sinusoid (black line).

To extract the amplitude $A_{DW}$, temporal period $\tau_{DW}$ and damping rate $\Gamma$ of the low frequency reflectivity modulations in metallic $(Sr_{1-x}La_x)_3Ir_2O_7$, we first subtracted a smoothly decaying background from the reflectivity transients and then fit the residual to a damped sinusoid of the form $A_{DW}e^{-\Gamma t}\sin(2\pi t/\tau_{DW} + \varphi)$, where $\varphi$ is a phase offset. Fits were constrained to the delay interval between 1.5 ps and 6 ps. The quality of fit for the $x = 5.8\%$ sample measured at 10 K is shown in Fig. S2, which is typical of all other temperatures and doping levels studied.

## S4. Fluence dependence of electronic mode in $(Sr_{1-x}La_x)_3Ir_2O_7$

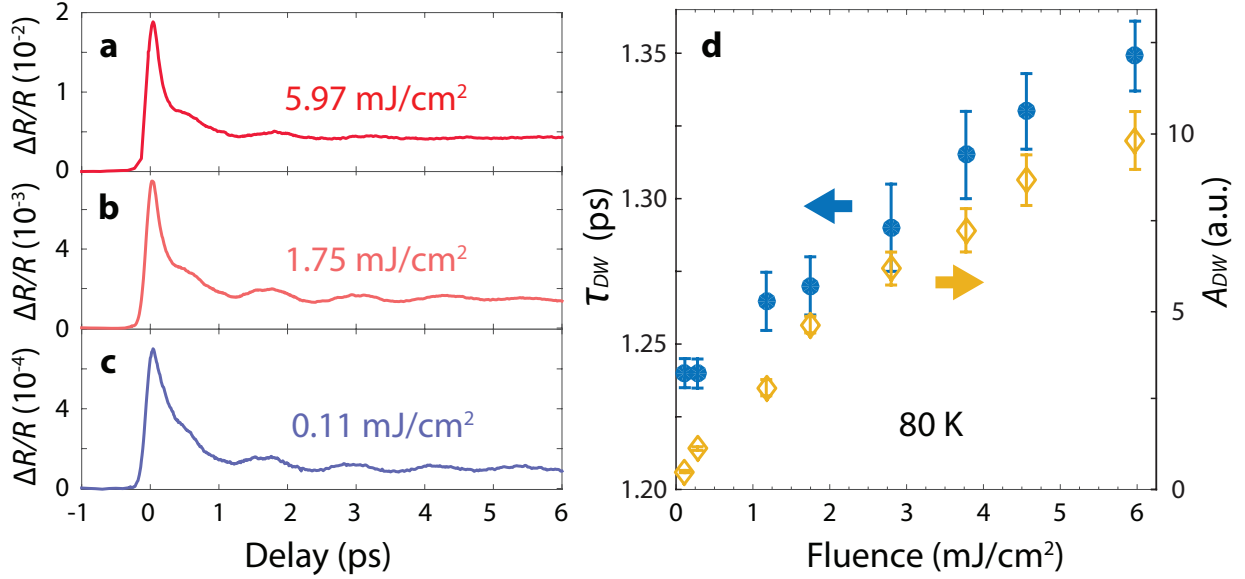

FIG. S3. Reflectivity transients of $(Sr_{1-x}La_x)_3Ir_2O_7$ with $x$ = 5.8 % measured at 80 K with a pump fluence of **a,** 5.97 mJ/cm$^2$, **b,** 1.75 mJ/cm$^2$ and **c,** 0.11 mJ/cm$^2$. **d,** The pump fluence dependence of the period $\tau_{DW}$ (filled circles) and amplitude $A_{DW}$ (open diamonds) of the reflectivity modulations observed in panels **a,** to **c,** extracted using the procedure described in section S3. Error bars represent the 95% confidence intervals of the fitted values.

To determine the pump fluence dependence of the low frequency reflectivity modulations in metallic $(Sr_{1-x}La_x)_3Ir_2O_7$, we measured the reflectivity transients of an $x$ = 5.8 % doped sample at a temperature far below $T_{DW}$ using a wide range of different pump fluences. As shown in Figs S3a-c, there is no qualitative change in the lineshape of the reflectivity transients over a roughly 50 fold increase in pump fluence. By fitting background subtracted reflectivity transients at different fluences to a damped sinusoid using the procedure described in section S3, we found that both the amplitude $A_{DW}$ and period $\tau_{DW}$ of the reflectivity modulations increase monotonically with pump fluence (Fig. S3d). This is consistent with amplitude oscillations of an electronic order parameter and contrasts with the pump fluence dependence of the $A_{1g}$ phonon mode, which grows in amplitude but does not noticeably shift in frequency as the pump fluence is raised.

## S5. La-doping dependence of $\tau_{DW}$

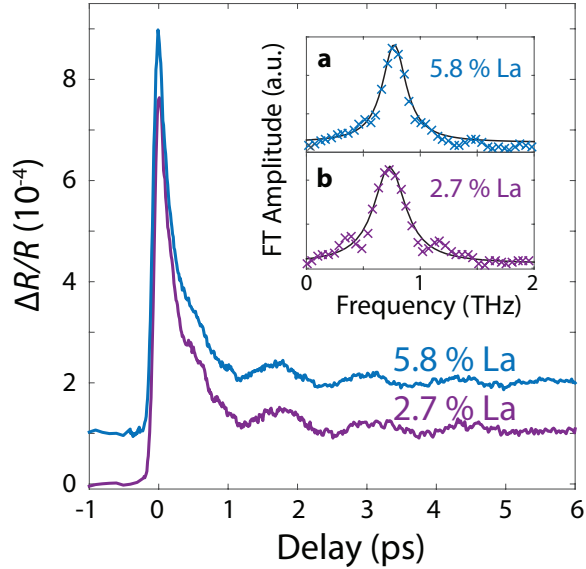

FIG. S4. Reflectivity transients of $x = 5.8$ % and $x = 2.7$ % doped $(Sr_{1-x}La_x)_3Ir_2O_7$ crystals measured at 100 K. Traces are vertically offset and scaled to similar magnitude for ease of comparison. Insets show the Fourier transforms of the background subtracted reflectivity transients of the **a,** $x = 5.8$ % and **b,** $x = 2.7$ % doped samples overlaid with fits to a Lorentzian function.

The temporal period $\tau_{DW}$ of the low frequency reflectivity modulations in metallic $(Sr_{1-x}La_x)_3Ir_2O_7$ is similar across all doping levels studied for any given temperature. As an example, Fig. S4 shows traces from metallic samples with the lowest ($x = 2.7$ %) and highest ($x = 5.8$ %) doping levels measured at 100 K. It is clear from the raw data alone that $\tau_{DW}$ is very similar for the two doping levels. We compared these traces more quantitatively by subtracting a decaying background as described in section S3 and then taking a Fourier transform. As shown in the insets of Fig. S4, the oscillation frequencies are the same for the two doping levels within our experimental resolution.

## S6. Characterization of structural distortion and $T_s$ in $(Sr_{0.93}Ca_{0.07})_3Ir_2O_7$

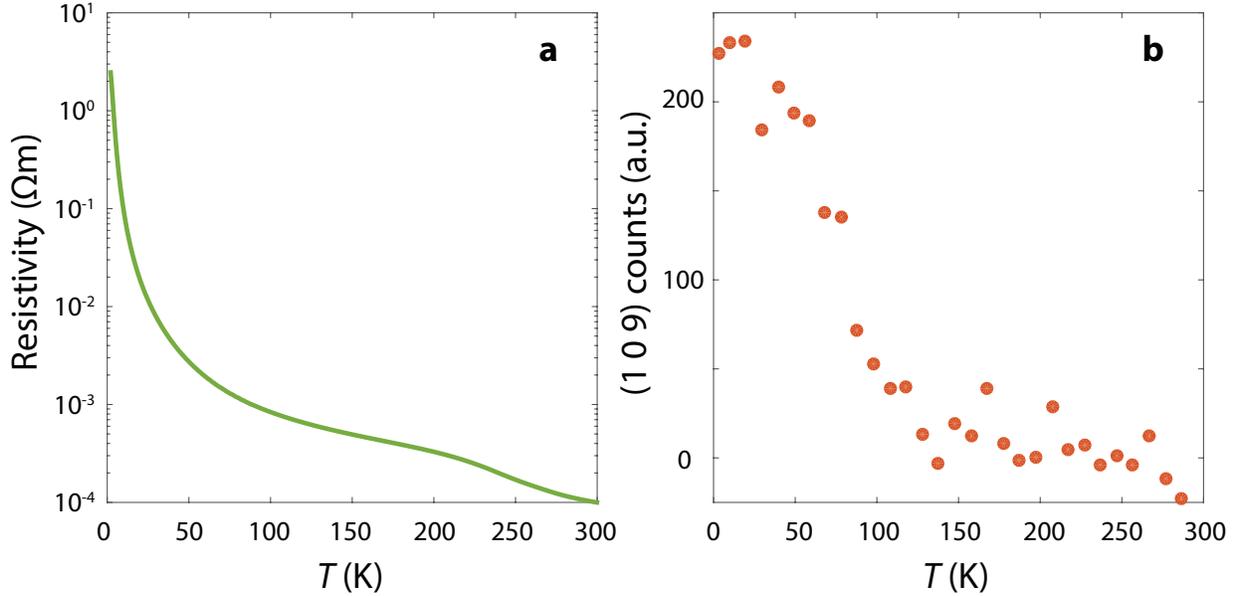

FIG. S5. Temperature dependence of **a,** the resistivity and **b,** the (1 0 9) structural Bragg peak measured by neutron diffraction of $(Sr_{0.93}Ca_{0.07})_3Ir_2O_7$.

The electrical properties of $(Sr_{0.93}Ca_{0.07})_3Ir_2O_7$ crystals from the same batch used in our study (see Fig. 4 of main text) were characterized by resistivity measurements, which were performed with a Lake Shore 370 AC Bridge using a four-probe lead configuration. A clear insulating behaviour that is similar to $Sr_3Ir_2O_7$ is observed (Fig. S5a), confirming the absence of a Fermi surface. The structural properties were characterized by neutron diffraction performed on the N5 triple-axis spectrometer at the Canadian Neutron Beam Centre at the Chalk River Laboratories. Like the case in metallic $(Sr_{1-x}La_x)_3Ir_2O_7$ samples, the weak structural distortion that occurs below $T_s$ in insulating $(Sr_{0.93}Ca_{0.07})_3Ir_2O_7$ samples appears in the form of very low intensity Bragg peaks at *Bbcb* space group forbidden ($h =$ odd, 0, $l =$ odd) positions. In particular, the array of peaks detected in the Ca-substituted system, namely (1, 0, 3), (1, 0, 7), (3, 0, 3) and (1, 0, 9), were also detected in the La-doped system, supporting the equivalence of the two structural distortions. The value of $T_s \sim 120$ K in $(Sr_{0.93}Ca_{0.07})_3Ir_2O_7$ was determined through the temperature dependence of the background subtracted (1 0 9) nuclear Bragg peak as shown in Fig. S5b.

These data sets collected from $(Sr_{0.93}Ca_{0.07})_3Ir_2O_7$ are reproduced from those shown in [2], where further details of the sample characterization can be found.

## S7. Magnon and spin-orbit exciton mode energies in $(Sr_{1-x}La_x)_3Ir_2O_7$

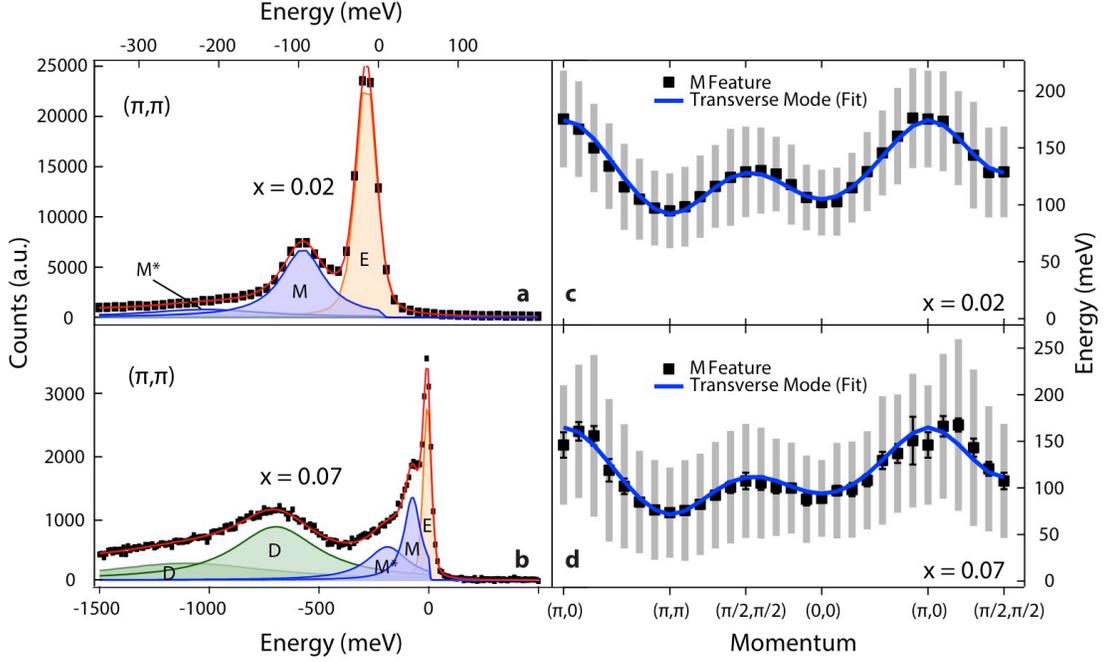

FIG. S6. (Reproduced from [3]). Ir $L_3$ edge (11.215 keV) RIXS energy scans collected at $T = 40$ K at a fixed momentum of $(\pi,\pi)$ for $(Sr_{1-x}La_x)_3Ir_2O_7$ samples with **a,** $x = 0.02$ and **b,** $x = 0.07$. Features labeled E, M, M* and D denote scattering from the elastic line, single magnon, multimagnon and spin-orbital exciton respectively. Dispersion of the M feature for **c,** $x = 0.02$ and **d,** $x = 0.07$ samples based on fitting to Lorentzian functions [3] are plotted as squares with accompanying fitting errors shown in gray. Solid lines denote fits to a bond operator model as described in [3].

Although the static long-range antiferromagnetic order observed in the parent compound $Sr_3Ir_2O_7$ is suppressed in the La-doped metallic samples, its remnant 2D magnetic fluctuations can still survive in the metallic regime. In order to study the possibility that the low energy mode observed using time-resolved optical reflectivity originates from coherent oscillations of such a fluctuating magnetic order parameter, we measured the dispersion of magnon excitations in $(Sr_{1-x}La_x)_3Ir_2O_7$ using resonant inelastic x-ray scattering (RIXS) at the Ir $L_3$ edge. A previous RIXS study of the parent compound $Sr_3Ir_2O_7$ [4] has shown that its excitation (magnon) spectrum is fully gapped, with the smallest gap size of $\Delta_{mag} \sim 90$ meV (= 22 THz) occurring at the $(\pi,\pi)$ point in the Brillouin zone. Upon La-doping, our RIXS data shows that these magnon excitations persist (see features labeled M in Fig. S6a,b) across the insulator-to-metal transition ($x_c \sim 0.02$) even though static long-range magnetic order is suppressed. By tracking the momentum dependence of this M feature in the RIXS spectra, we obtained the magnon dispersion relations (Fig. S6c,d) showing that the gap at ($\pi,\pi$) in the $x = 0.02$ ($\Delta_{mag} \sim 90$ meV = 22 THz) and $x = 0.07$ ($\Delta_{mag} \sim 70$ meV = 17 THz) samples stays at a value comparable to that observed in the parent compound. Since the frequency of the mode observed in our time-resolved optical reflectivity measurements (~ 1 THz) is far lower than the lowest energy magnon across all La-doping values measured, we rule out a magnon origin of this mode.

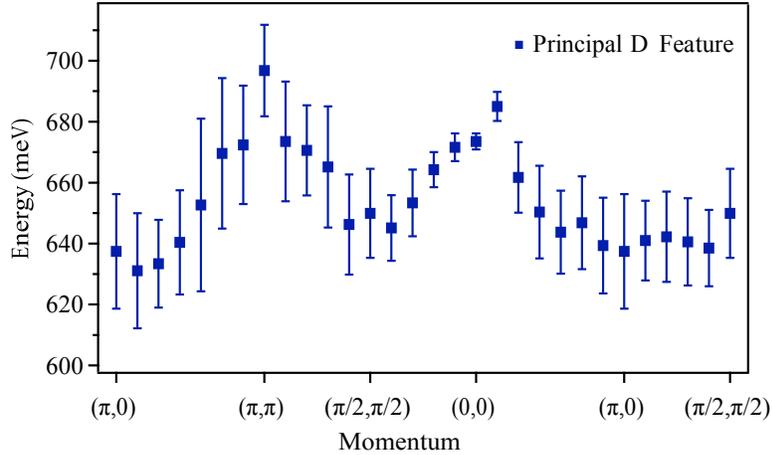

FIG. S7. Dispersion of the spin-orbit exciton mode in the metallic $x = 0.07$ sample (see feature labeled D in Fig. S6**b**) obtained by fitting the RIXS data to Lorentzian functions [3] are plotted as squares with accompanying fitting errors.

We also studied the possibility that the mode observed by time-resolved optical reflectivity originates from a charge neutral spin-orbit exciton mode (a bound state of a $J_{eff} = 3/2$ hole and a $J_{eff} = 1/2$ electron) that has very recently been observed by RIXS in parent $Sr_3Ir_2O_7$ [5]. Optical conductivity measurements [6] have shown that the broad $J_{eff} = 3/2$ and $J_{eff} = 1/2$ manifolds in $Sr_3Ir_2O_7$ are separated in energy by approximately 0.8 eV and therefore $J_{eff} = 3/2$ to $J_{eff} = 1/2$ excitations are energetically accessible by our pump photons ($\hbar\omega = 1.5$ eV). Our RIXS measurements have also detected the spin-orbit exciton in metallic La-doped $Sr_3Ir_2O_7$ samples at this energy scale (see principal feature labeled D in Fig. S6b) and its complete dispersion relation is shown Fig. S7. As is the case with the magnon spectrum, the spin-orbit exciton mode is fully gapped with a gap energy scale of around 650 meV =

157 THz. Since this again far exceeds the frequency of the mode observed in time-resolved optical reflectivity, we can also rule this out as a possible origin.

---